\documentclass[aps,pre,twocolumn,groupedaddress,10pt]{revtex4-1}
\usepackage[dvips]{graphicx}
\usepackage{amsmath}
\usepackage{amssymb}
\usepackage{color}
\usepackage{bm}

\newcommand{\commut}[2]{\left[#1,#2\right]}
\newcommand{\Trace}[1]{\textrm{Tr}\left(#1\right)}
\newcommand{\ket}[1]{|#1\rangle}
\newcommand{\bra}[1]{\langle#1|}
\newcommand{\brket}[3]{\langle#1|#2|#3\rangle}

\newcommand{\trm}[1]{\textrm{#1}}

\begin{document}


\title{Quantum Recurrences in a One-Dimensional Gas of Impenetrable Bosons}


\author{E. Solano-Carrillo}
\affiliation{Department of Physics, Columbia University, New York, NY 10027, USA.}



\begin{abstract}

It is well-known that a dilute one-dimensional (1D) gas of bosons with infinitely strong repulsive interactions behaves like a gas of \emph{free} fermions. Just as with conduction electrons in metals, we consider a single-particle picture of the resulting dynamics, when the gas is isolated by enclosing it into a box with hard walls and preparing it in a special initial state. We show, by solving the nonstationary problem of a free particle in a 1D hard-wall box, that the single-particle state recurs in time, signaling the intuitively expected back-and-forth motion of a free particle moving in a confined space. Under suitable conditions, the state of the whole gas can then be made to recur if all the particles are put in the same initial momentum superposition. We introduce this problem here as a modern instance of the discussions giving rise to the famous recurrence paradox in statistical mechanics: on one hand, our results may be used to develop a poor man's interpretation of the recurrence of the initial state observed [T. Kinoshita et al, Nature \textbf{440}, 900 (2006)] in trapped 1D Bose gases of cold atoms, for which our estimated recurrence time is in fair agreement with the period of the oscillations observed; but this experiment, on the other hand, has been substantially influential on the belief that an isolated quantum many-body system can equilibrate as a consequence of its own unitary nonequilibrium dynamics. Some ideas regarding the latter are discussed.
\end{abstract}


\maketitle
The mechanism for the approach to statistical equilibrium of isolated many-body systems is one of the oldest problems in science which is still awaiting a satisfactory solution. It first grew in popularity with a tough debate \cite{Steckline} in the late 19th century starred by Boltzmann, who thought about molecular chaos as an ingredient for the equilibration process, and Zermelo who, using a theorem of Poincar\'e, argued that a given initial state would always recur in time, raising doubts of any equilibration at all. With modern experimental techniques able to reach a high degree of isolation, the problem of the nonequilibrium dynamics of isolated quantum many-body systems can now be addressed as never before, specially in cold-atom systems \cite{Kinoshita,Hofferberth,Trotzky,Gring,Polkovnikov}. A remarkable outcome is given by the  \emph{persistent} oscillations of the initial state observed by Kinoshita, Wenger and Weiss (KWW) \cite{Kinoshita} in trapped 1D Bose gases of cold atoms, where equilibration is not observed even after thousands of atomic collisions. The long-time behavior of this system is believed \cite{Rigol} to be an unsual kind of equilibrium, which carries a good memory of the initial state through conserved quantities, and described by a generalized Gibbs ensemble (GGE).

We recall that a crucial property to reach equilibration in (infinite) quantum systems with a continuous energy spectrum, as typical condensed matter systems in the thermodynamic limit are usually treated, has been recognized long time ago by Van Hove \cite{vHove2,vHove,Fujita} and marked the foundations of modern nonequilibrium statistical mechanics \cite{Zwanzig,vVliet1,vVliet2,vVliet3}. That is, when the approach to equilibrium can be attributed to a perturbation term in the Hamiltonian that is able to produce self-energy effects, as is the case, e.g., of phonon-phonon interaction in the theory of heat conduction in crystals or magnon-magnon and magnon-phonon interaction in ferromagnetic relaxation phenomena, then starting from initial statistical states which are diagonal in the energy representation of the unperturbed system, the expectation value of all diagonal operators, which happen to be the slowly-varying or thermodynamic observables of the quantum theory (in the ideal limit of single-level energy resolution \cite{vKampen}), can be rigorously proved to tend to their microcanonical values in the long-time limit. 

For finite isolated quantum many-body systems, which have discrete energy spectrum, there is a natural reason to believe that equilibration can also take place, at least partially and in a definite time window \cite{Yukalov}, leading to a GGE \cite{Rigol} accounting for the full set of conserved quantities in integrable systems, or again to a microcanonical ensemble for nonintegrable systems \cite{Rigol2,Rigol3}. Loosely speaking, this is due to dephasing: the idea that the probability of constructive interference among the stationary states comprising a given initial state, which gives periodic behavior in few-body systems, decreases with time under the effect of the perturbation, as more degrees of freedom (e.g. particles) are in the system. The time evolution in the long-time limit is then expected to be described by a diagonal ensemble, where any reference to the quantum coherences disappears.

Although dephasing or the random-phase assumption for wave-function interference, is expected to be true for an overwhelming majority of initial states in infinite systems \cite{vHove},
there are many-body examples, such as the collapse and revival of the population inversion in the one-atom maser \cite{Shore}, of the matter-wave field of a Bose-Einstein condensate \cite{Greiner}, or the spin echoes in pulsed NMR experiments \cite{Hahn}, where constructive interference appears at a time when the system is apparently observed in a steady state, and therefore the dephasing hypothesis should be taken with caution when applied to systems with a large, but finite, number of degrees of freedom. In this paper, we investigate a different alternative for the dynamics of a finite isolated quantum many-body system, which is the recurrence of its initial state as opposed to equilibration, taking the KWW observations as a test experiment with an observable recurrence time.

In the geometrical picture offered by the Liouville representation of quantum mechanics \cite{Fano}, we can see all operators acting on Hilbert space as vectors, with an inner product between any two operators $A$ and $B$ defined as $(A,B)=\trm{Tr}\,(AB^{\dagger})$.
The general idea of quantum recurrences exploited here comes from the fact that the \emph{unitary} evolution of an isolated quantum system conserves the norm $(\rho,\rho)=\Trace{\rho^2}$ and then the density matrix, seen as a vector in Liouville space, must be rotating about some generalized axes. The situation we will encounter is then reminiscent of the classical problem of finding the principal axes about which the general complicated rotational motion of a 3D rigid body can be seen as elementary orthogonal rotations. As is well-known, these axes are closely connected with the symmetries of the body and then, following this analogy, we should look for an operator basis or ``coordinates'', $U_{\alpha}$, of Liouville space connected to the relevant symmetry properties of the system. The quantum interference of the corresponding elementary oscillations in our test case will be shown to produce the recurrent behavior.

We recall that a complete orthonormal operator basis, $U_{\alpha}$, with the identity matrix belonging to the set, can always be constructed out of the generators of infinitesimal unitary transformations in the Hilbert space of any quantum-mechanical system. These operators satisfy the Lie-algebraic relations $i\,[U_{\alpha},U_{\beta}^{\dagger}]=\sum_{\gamma}c_{\alpha\beta}^{\gamma}U_{\gamma}$ where the so-called structure constants, $c_{\alpha\beta}^{\gamma}$, are antisymmetric under the interchange of lower indices. The completeness property allows any operator in Liouville space to be expanded as $A=\sum_{\alpha} (A,U_{\alpha}^{\dagger})\,U_{\alpha}^{\dagger}\equiv \sum_{\alpha}A_{\alpha}U_{\alpha}^{\dagger}$. In this geometrical picture, the Liouville-von Neumann equation for the evolution of the density matrix reads \cite{Fano}
\begin{equation}\label{drt}
 \dfrac{\partial\bm{\rho}}{\partial t}=\,\Omega\,\cdot\bm{\rho},
\end{equation}
where $\Omega_{\alpha\beta}\equiv \sum_{\gamma}c_{\alpha\beta}^{\gamma}H_{\gamma}$, with $H_{\gamma}$ the components of the Hamiltonian along $U_{\gamma}^{\dagger}$, and $\bm{\rho}$ is the vector with components $\rho_{\alpha}$ along $U_{\alpha}^{\dagger}$. Due to the antisymmetry of the structure constants, it is easy to show that $(\partial\bm{\rho}/\partial t,\bm{\rho})=0$ and then the action of $\Omega$ on $\bm{\rho}$ is an orthogonal transformation which continually ``rotates'' the density matrix in Liouville space.

The simplest example of this formalism is a system of noninteracting two-level systems. The SU(2) symmetry of these systems suggests the single-particle operator basis  $\left\lbrace\hat{1},\sigma_x,\sigma_y,\sigma_z\right\rbrace$, with respect to which the rotation of the density matrix in Liouville space leads to well-known recurrent phenomena such as the Larmor precession of magnetic moments in a magnetic field \cite{Fano2}, or the Rabi oscillations of two-level atoms (dipoles) in a microwave field \cite{Feynman}. Higher-dimensional examples, still with few degrees of freedom, can be developed \cite{Fano,Fano3} by constructing the corresponding symmetrical operator basis. We contemplate here the application of this formalism to a system with an almost infinite number of degrees of freedom. It is interesting by itself for pedagogical reasons since it constitutes, in the end, a sophisticated \emph{nonstationary} version of the textbook quantum-mechanical problem of a particle in a box undergoing a back-and-forth motion.

\begin{figure}[t]
 \centering
\includegraphics[scale=0.4]{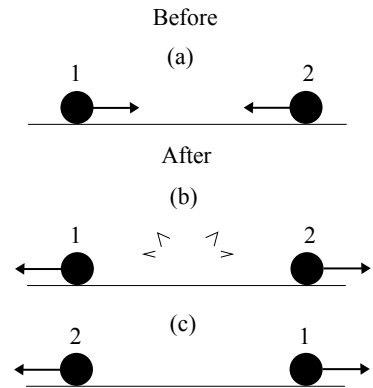}
\caption{One dimensional elastic collision of point particles with equal masses. (a) The particles are approaching each other with equal speeds before the collision. (b) The only possible outcome after the collision if the particles are distinguishable classical objects. (c) Allowed outcome if the particles are quantum objects. In this case there is perfect transmission of one particle through the other and the situation is physically indistinguishable from the perfect reflection in (b).\label{coll}}
\end{figure}

We consider an ensemble of spatially \emph{non-overlaping} point particles with equal masses confined in a 1D box, and split up into two groups, the particles in different groups differing by the direction of their momentum $|k|$. We note that, classically, for a pair of particles of different groups undergoing a collision, the result of their contact interaction is the perfect transmission of the momentum of one of the particles to the other, as shown in Fig. \ref{coll}. For indistinguishable quantum particles, however, tunneling may occur after the collision, and the outcome cannot be physically distinguished from a perfect reflection. We can then treat these particles as ``ghosts''\cite{Kinoshita} perfectly transmiting through each other and concentrate on the dynamics of just one of them.

The system described above can be used as a poor man's version of a trapped 1D system of interacting bosons, as in the KWW experiment, in which the harmonic longitudinal confinement is instead replaced by a flat-bottom-like potential resembling the confinement to a box. In the dilute limit of infinitely strong repulsive interactions (a.k.a. gas of impenetrable bosons or Tonks-Girardeau gas) the single-particle spacing is of the order of the spread of the single-particle wave functions \cite{Kinoshita2}, and then the particles in the same group are essentially non-overlapping. Since in this regime the particles behave as \emph{free} fermions \cite{Dunjko}, it is then justified to use the single-particle picture refered to above, just as it is justified to use a single-particle picture to study the nonequilibrium dynamics of conduction electrons in metals.

The problem is therefore that of a free particle in a 1D box. Note that a general way to construct a complete unitary and orthonormal operator basis for any quantum system with Hilbert space dimension $N$ was given by Schwinger \cite{Schwinger} in terms of complementary pairs of shift operators. In the continuum case, which is approached as a limit \cite{Schwinger2}, the basis vectors in Liouville space are elements of the Heisenberg group \cite{Patra}. We take the Schwinger basis $U_{qp}=e^{ip\hat{q}}e^{iq\hat{p}}$, where $\hat{q}$ and $\hat{p}$ are, respectively, the position and momentum operators of the particle, and where $q$ and $p$ are \emph{values} which change (with $a$ the lattice constant) in steps $\Delta q=a,\,\Delta p=2\pi/Na$. In these coordinates for the Liouville space, the state is represented by the components, $\rho_{qp}=(U_{qp},\rho)$, of the density matrix along the unit vectors. 

The confinement of the particle to the 1D (flat-bottom) box, which is centered at the origin and of length $L=(N-1)a$, is modeled by repulsive delta potentials,  $V(\hat{q})=(gL/2)\,\bigl[\delta(\hat{q}-L/2)+\delta(\hat{q}+L/2)\bigr]$, at the box edges (impenetrable walls). In the following, we consider the continuum limit, $N (\textrm{odd})\rightarrow\infty$, $a\rightarrow0$, with $L$ finite. Then, in units of $2m=1$, the component of the Hamiltonian of the ``free'' particle  along the unit vector $U_{qp}^{\dagger}$ is given by $H_{qp}=\varepsilon_{q}\delta(p)+v_p\delta(q)$, where we have defined the Fourier transform $\varepsilon_q=(1/2\pi)\int dp'e^{iqp'}p'^2$, and $v_p=gL\cos(p\,L/2)$. Using the structure constants of the Heisenberg group,  $c_{qp,q'p'}^{rs}=i[e^{-ip'(q-q')}-e^{-iq'(p-p')}]\delta_{r,q-q'}\delta_{s,p-p'}$, we then find that \eqref{drt} is equivalent in this case to the spinor equation (see Appendix \ref{apb})

\begin{equation}\label{dv}
 \dfrac{\partial}{\partial t} 
\begin{pmatrix}
 \tilde{\rho}_{kp}\\
\tilde{\rho}_{-k,p}^{*}
\end{pmatrix}
= i
\begin{pmatrix}
 \tilde{\varepsilon}_{p-k}-\tilde{\varepsilon}_{-k} & 0\\
0 & \tilde{\varepsilon}_k-\tilde{\varepsilon}_{k+p}
\end{pmatrix}
\begin{pmatrix}
 \tilde{\rho}_{kp}\\
\tilde{\rho}_{-k,p}^{*}
\end{pmatrix},
\end{equation}
where the tilde denotes an inverse Fourier transform in the $q$ index, which diagonalizes $\Omega$ and then represents the ultimate transformation to the principal axes. Here, $\tilde{\varepsilon}_k=k^2$ is the \emph{bare} energy of a particle with momentum $k$, and $\tilde{\rho}_{kp}=\brket{k}{\,\rho\,}{k+p}$ is the transition amplitude, $\ket{k}\rightarrow \ket{k+p}$, between momentum eigenstates. Note that, any reference to the properties of the walls is translated to the boundary condition quantizing the momentum in integer factors of $\pi/L$. 

\begin{figure}[t]
 \centering
\includegraphics[scale=0.4]{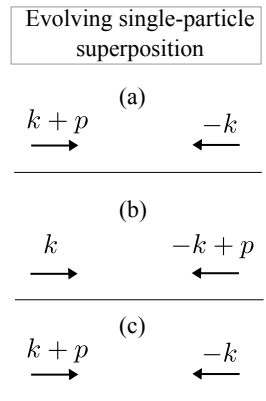}
\caption{Time evolution of an initial single-particle superposition. (a) Initially the particle is described by a superposition of two waves ``freely'' propagating in opposite directions and with the magnitude of their momentum slightly differing by an amount $p$. (b) After half a recurrence period, the component waves have perfectly reflected off the walls. (c) A subsequent wall reflection brings the single-particle superposition to the initial configuration, and an exact recurrence is completed.\label{mt}}
\end{figure}

The results can now be used to explain the KWW experiment in the strong-coupling regime if we interpret $p$ as a slight momentum imbalance between the atoms in the two groups, arising due to imperfection in their preparation.  As can be readily observed from Fig. \ref{mt}, we get oscillations of the states $\ket{k}\leftrightarrow\ket{k+p}$ embodied in $\tilde{\rho}_{kp}$, which take place at the frequency $\omega_2=\tilde{\varepsilon}_{p-k}-\tilde{\varepsilon}_{-k}$ and, similarly, of the states $\ket{-k+p}\leftrightarrow\ket{-k}$ embodied in $\tilde{\rho}_{-k,p}^{*}$ and taking place at the frequency $\omega_1=\tilde{\varepsilon}_{k}-\tilde{\varepsilon}_{k+p}$. The quantum interference of these modes give oscillations with the average frequency $|\omega_2+\omega_1|/2=2kp$, from which we identify the recurrence time, which in an \emph{ideal} experiment ($p=0$), interpreted as the limit $|k|\gg p$ (i.e. $p=p_{\textrm{min}}=\pi/L$ the minimum nonzero wave vector in the reciprocal lattice) gives, $\tau_k=2\pi/(2kp)_{p=p_{\textrm{min}}}$, or
\begin{equation}\label{tk} 
 \tau_k=\dfrac{2mL}{\hbar k}=\dfrac{2mL^2}{\pi\hbar n_k},
\end{equation}
where $n_k$ is an integer representing the position of $k$ in reciprocal space, that is $k=n_k \pi/L$. The period of these recurrences corresponds, classically, to the time it takes for the back-and-forth motion of a free particle in a 1D box, if we take $v_k=\hbar k/m$ as the classical velocity.

That the result we have established from a single-particle picture is valid in the Tonks-Girardeau limit of a 1D system of impenetrable bosons is supported by the results obtained recently by Kaminishi et al \cite{Kaminishi} who, by calculating the square amplitude between an initial state and the time-evolved state, and observing its periodicity, derived rigorously our $L^2$-law for the recurrence time in the \emph{free-fermionic} and free-bosonic limit of the Lieb-Liniger model, independent of the initial state. They made an estimate of the order of $10$ ms for cold atoms confined in one dimension of $10\,\mu$m, in a superposition of Lieb's type II (one-hole) excitations, and posed the challenge for its observation. This is of the same order of magnitude that we estimate for the initial state of the KWW experiment in the strong-coupling regime (see Appendix \ref{kww}).

In the KWW experiment, the atoms go back-and-forth without noticeably equilibrating even after thousands of collisions. A slow relaxation of the initial momentum distribution is observed, attributed to dephasing of the oscillating atoms due to trap anharmonicities, although the effect of the boundary conditions, revealing the lack of perfect isolation is believed here to play an important role. In fact, we can show within our picture (see Appendix \ref{apb}) that a slight deviation from the hard-wall boundary condition, meaning a small but nonzero probability for the particles to leave the box, produces \emph{decoherence}, manifested as a slow decay with time of the off-diagonal elements of the single-particle density matrix, $\tilde{\rho}_{kp}$, superimposed to the oscillations already discussed. This source of decoherence is actually measured in the KWW experiment, as atomic losses (besides others such as heating effects) and points to the fact that a relaxation towards a steady state in the nonequilibrium dynamics of a quantum many-body system requires, in some degree, the contact with an environment.

The possibility for an isolated quantum system to approach equilibrium without the need of an environment is still there but requires a more explicit theoretical demonstration of how exactly this can take place. To illustrate this, let us consider isolated integrable systems for which the long-time expectation value of local operators has been conjectured \cite{Rigol} to be described by a GGE, which is represented by $\rho_{\textrm{GGE}}\propto \exp(-\sum_i \alpha_i I_i)$, where the $I_i$ are the local conserved quantities, and the $\alpha_i$ are Lagrange multipliers chosen so as to ensure that these conserved quantities remain constant over time. Hard-core bosons in a 1D finite lattice fall in this cathegory and constitute a model studied numerically in \cite{Rigol} to emulate the conditions in the KWW experiment. For a satisfactory answer to the question posed there of whether the system relaxes to an equilibrium state (being described by GGE), it must then be proved that the unitary dynamics take the initial density matrix along a path whose asymptotics coincides with $\rho_{\textrm{GGE}}$.

A proof that a steady-state density matrix of an exponential form (as the GGE is) can arise from the unitary dynamics of a general isolated quantum system was given sometime ago by Hershfield \cite{Hershfield}, assuming that an unspecified physical relaxation process causes correlation functions to decay at long times. This assumption, which is hard to imagine without the presence of an environment (or wrong boundary conditions playing the role of it), can be avoided for the steady state in some transport problems by making use of the ``open system limit'' \cite{Dutt}, which however requires going to the thermodynamic limit. It is then challenging to show that the GGE is the natural fate for the unitary time evolution of the density matrix in a finite system, as considered here. 

In summary, we have shown that in a finite isolated quantum many-body system such as a 1D gas of impenetrable bosons enclosed in a hard-wall box, where a single-particle picture of the whole dynamics is possible due to fermionization and the preparation of all the particles in the same initial superposition, that the initial state of the gas inevitably recurs as a consequence of the unitary evolution, in contrast to the possibility that the gas equilibrates. The recurrence time is proportional to $L^2$, with $L$ the length of the box, and by using our model as poor man's version of the conditions in the KWW experiment in the strong-coupling regime, this recurrence time is found to be in fair agreement with the oscillations of the initial state observed in that experiment in the mentioned regime. The possibility of having trajectories for the density matrix of an isolated quantum system, which in the long-time limit tend to a time-independent exponential form, as the GGE, will be investigated elsewhere.

\ \vspace{0.3cm}

\begin{acknowledgments}
 I am grateful to D. Weiss for clarifying some details of the KWW experiment and T. Deguchi, J.-S. Caux, R. van den Berg, and L. F. Santos for useful comments in the early stage of this work. Support from the Fulbright-Colciencias fellowship, from the Columbia GSAS faculty fellowship, from ICAM and from Prof. Andy Millis is acknowledged.
\end{acknowledgments}
\appendix
\section{Estimating the recurrence time in the KWW experiment.}\label{kww} In the KWW experiment \cite{Kinoshita}, there is an ensemble of thousands parallel (and non-interacting) 1D Bose gases, with a number of $^{87}$Rb atoms ranging from 40 to 250 in each tube. These are put, in an initial time, in a momentum superposition with $\hbar k_{e}$ to the right and $\hbar k_e$ to the left, with $k_e\equiv 2k$ determined from the total atomic collision energy $8(\hbar k)^2/2m=0.45\,\hbar\omega_r$, where $\omega_r$ is the lowest transverse excitation frequency ($\omega_r/2\pi=67$ kHz). An ``effective'' length, $L$, for the statistical ensemble of parallel Bose gases can be determined from the weighted average of the 1D coupling strength in each tube, $\gamma_0=|2/a_{\textrm{1D}}n_{\textrm{1D}}|$, and the 1D density $n_{\textrm{1D}}=N_{\textrm{tube}}/L$, where $N_{\textrm{tube}}$ is the weighted average of atoms per tube, $|a_{\textrm{1D}}|\approx a_r^2/2a$ is the 1D scattering length, with $a_r=41.5$ nm the transverse oscillator width, and $a=5.3$ nm the 3D scattering length. By taking, say $N_{\textrm{tube}}=211$, and the value $\gamma_0=4$ reported in the experiment for the strong-coupling regime, we get an effective length of $L\simeq70\,\mu$m, whereby $n_{k_e}\simeq353\gg1$, as required from the condition $|k_e|\gg p$. The corresponding recurrence time, from \eqref{tk}, is $\tau_{k_e}\simeq 12$ ms, which is to be compared with the observed value $\tau=34$ ms.
\section{Derivations}\label{apb}
\noindent \textbf{Conventions:} We note that, originally, Schwinger used \cite{Schwinger2} a convention in which $\Delta q=\Delta p =\sqrt{2\pi/N}$ which is equivalent to the symmetric direct and inverse Fourier transforms. Here we use the standard convention for direct and reciprocal lattices, namely, $\Delta q=a$ (with $a$ the lattice constant) and $\Delta p=2\pi/Na$, which conserve the phase-space volume $\Delta q \Delta p$, and we consider the limit $a\rightarrow0$, $N\rightarrow\infty$ with $(N-1)a=L$ fixed. 
\subsection{Hamiltonian and density matrix in Liouville space}
First we need to project all operators along the Schwinger coordinates, $U_{qp}=e^{ip\hat{q}}e^{iq\hat{p}}$, which can be shown to constitute a complete othonormal operator basis. For the Hamiltonian, $H=\hat{p}^2+V(\hat{q})$, we need to calculate $H_{qp}=(U_{qp},H)=\textrm{Tr}\,[\,\hat{p}^2+V(\hat{q})\,]e^{ip\hat{q}}e^{iq\hat{p}}$, that is
\begin{widetext}
\begin{equation}\label{ham}
\begin{split}
  H_{qp}&=\int \dfrac{dp'}{2\pi}\brket{p'}{\,\hat{p}^2\,e^{ip\hat{q}}e^{iq\hat{p}}}{p'}+\int dq'\brket{q'}{\,V(\hat{q})\,e^{ip\hat{q}}e^{iq\hat{p}}}{q'},\\
&=\int \dfrac{dp'}{2\pi} e^{iqp'}p'^2\delta(p'-(p'+p))+\int dq' e^{ipq'}V(q')\delta(q'-(q'+q)),\\
&=\varepsilon_q\,\delta(p)+v_p\,\delta(q),
 \end{split}
\end{equation}
\end{widetext}
where we have defined $\varepsilon_q=(2\pi)^{-1}\int dp' e^{iqp'}p'^2$ and, with $V(q')=(gL/2)[\delta(q'-L/2)+\delta(q'+L/2)]$, we easily get $v_p=gL\cos(pL/2)$. Note that we have used the shift operator properties: $e^{ip\hat{q}}\,\ket{p'}=\ket{p'+p}$ and $\bra{q'}\,e^{iq\hat{p}}=\bra{q'+q}$. For the density matrix we need to calculate $\rho_{qp}=(U_{qp},\rho)=\textrm{Tr}\,\rho\,e^{ip\hat{q}}e^{iq\hat{p}}$, that is
\begin{equation}\label{dens}
 \begin{split}
  \rho_{qp}=\int \dfrac{dp'}{2\pi}\brket{p'}{\,\rho\,e^{ip\hat{q}}e^{iq\hat{p}}\,}{p'}=\int \dfrac{dp'}{2\pi}e^{iqp'}\brket{p'}{\,\rho\,}{p'+p}.
 \end{split}
\end{equation}
\subsection{Structure constants of the Heisenberg group}
The structure constants of the Heisenberg group are obtained from $i\ [U_{\alpha},U_{\beta}^{\dagger}]=\sum_{\gamma}c_{\alpha\beta}^{\gamma}U_{\gamma}$, by repeated use of the Baker-Campbell-Hausdorff formula in the form $e^{\hat{x}}e^{\hat{y}}=e^{\hat{x}+\hat{y}+\commut{\hat{x}}{\hat{y}}/2}$ which holds whenever $\commut{\hat{x}}{\hat{y}}$ is proportional to the identity matrix. This leads, with $\commut{\hat{q}}{\hat{p}}=i$, to the more convenient form  $$U_{qp}=e^{i(p\hat{q}+q\hat{p}-qp/2)},$$ and hence
\begin{multline*}
\commut{U_{q'p'}}{U_{qp}^{\dagger}}=\bigl[e^{-ip(q'-q)}-e^{-iq(p'-p)}\bigr]\\ \times e^{i[(p'-p)\hat{q}+(q'-q)\hat{p}-(q'-q)(p'-p)/2]}.
\end{multline*}
Rewriting this as $i\commut{U_{q'p'}}{U_{qp}^{\dagger}}=\sum_{r,s}c_{q'p',qp}^{rs}U_{rs}$ we get
\begin{equation}\label{cs}
 c_{q'p',qp}^{rs}=i\bigl[e^{-ip(q'-q)}-e^{-iq(p'-p)}\bigr]\delta_{r,q'-q}\delta_{s,p'-p}.
\end{equation}
A mixed notation with sums and integrals will be kept to remind that the values of $q$ and $p$, although possibly very large, remain \emph{finite}  \cite{Schwinger2}.
\subsection{Combining all in the Fano equation}
Using \eqref{ham} to \eqref{cs} we can express the Liouville-Von Neumann equation for the density matrix in Liouville space \cite{Fano} as
\begin{widetext}
\begin{equation}\label{feq}
 \begin{split}
  \dfrac{\partial \rho_{qp}}{\partial t}&= \int\int dq' dp'\,\Omega_{qp,q'p'}\rho_{q'p'}=\sum_{rs}\int\int dq' dp'c_{qp,q'p'}^{rs}H_{rs}\rho_{q'p'},\\
&=i\sum_{rs}\int\int dq' dp'\bigl[e^{-ip'(q-q')}-e^{-iq'(p-p')}\bigr]\delta_{r,q-q'}\delta_{s,p-p'}\,[\varepsilon_r\,\delta(s)+v_s\,\delta(r)]\,\rho_{q'p'},\\
&=i\int\int dq' dp'\bigl[e^{-ip'(q-q')}-e^{-iq'(p-p')}\bigr]\,[\varepsilon_{q-q'}\,\delta(p-p')+v_{p-p'}\,\delta(q-q')]\,\rho_{q'p'},\\
&=i\int dq'\bigl[e^{-ip(q-q')}-1\bigr]\varepsilon_{q-q'}\rho_{q'p}+i\int dp'\bigl[1-e^{-iq(p-p')}\bigr]v_{p-p'}\rho_{qp'}.
 \end{split}
\end{equation}
\end{widetext}
\subsection{Transformation to principal axes}
We now perform an inverse Fourier transformation over the $q$ index. Define  
\begin{equation}
\tilde{\rho}_{kp}=\int dq\,e^{-ikq}\rho_{qp},\hspace{0.5cm}\textrm{and}\hspace{0.5cm}\tilde{\varepsilon}_k=\int dq\,e^{-ikq}\varepsilon_q.
\end{equation}
With these we can deconvolve \eqref{feq} and obtain our main result
\begin{equation}\label{dr}
\dfrac{\partial \tilde{\rho}_{kp}}{\partial t}=i(\tilde{\varepsilon}_{k-p}-\tilde{\varepsilon}_k)\,\tilde{\rho}_{kp}+i\int dp'v_{p'}\bigl(\tilde{\rho}_{k,p-p'}-\tilde{\rho}_{k-p',p-p'}\bigr).
\end{equation}
Note that, from the definitions, we easily get
\begin{equation}
\tilde{\rho}_{kp}=\brket{k}{\,\rho\,}{k+p},\hspace{0.5cm}\textrm{and}\hspace{0.5cm}\tilde{\varepsilon}_k=k^2,
\end{equation}
which are the transition amplitudes between the momentum eigenstates $\ket{k}\rightarrow\ket{k+p}$, and the bare-particle energies, respectively.
\section{Boundary conditions}
\subsection{Hard walls}
In the textbook problem of a free particle in a symmetric box, i.e. with the origin at the center, the hard-wall condition is imposed by the vanishing of the wavefunction at the box walls. The stationary states (which are also the momentum eigenstates) then have either even parity with $k=n_{\textrm{odd}}\pi/L$, and \emph{nonzero} probability for the particle to be at the center, or odd parity with $k=n_{\textrm{even}}\pi/L$ and \emph{zero} probability for the particle to be at the center. In our lattice problem, with $N$ odd (as in Schwinger's treatment), the center of the box is a lattice site, and then the particle is likely to hope to it. This means that all our momentum eigenstates must have the same (even) parity, a reason why $\Delta p=2\pi/L$ and not $\Delta p=\pi/L$. It is then easy to see that, imposing vanishing transition amplitudes between momentum eigenstates with different parities makes the second term in \eqref{dr} vanish. 

Alternatively, we can impose the hard-wall boundary condition by saying that the probability of the transition $\ket{k}\rightarrow\ket{k+p}$ does not depend on $k$. This can be written as $\brket{k+k'}{\,\rho\,}{k+k'+p}=e^{-ik'L}\brket{k}{\,\rho\,}{k+p}$, which implies
\begin{equation}\label{hw}
\tilde{\rho}_{k-p',p-p'}=e^{ip'L}\,\tilde{\rho}_{k,p-p'}.
\end{equation}
With this, the second term in \eqref{dr} is proportional to $\sum_{p'}\Delta p'\sin(p'L)e^{ip'L/2}\tilde{\rho}_{k,p-p'}=0$, vanishing due to momentum quantization, and then \eqref{dv} in the main article follows.
\subsection{Slightly penetrable walls}
Motivated by \eqref{hw} we can imagine a boundary condition which produces decaying coherences. Consider, for example,
\begin{equation}
 \tilde{\rho}_{k-p',p-p'}=\bigl[1-iL^{-1}\delta(p')(\tilde{\varepsilon}_k/g)^2\bigr]\,\tilde{\rho}_{k,p-p'}.
\end{equation}
In this case, the second term in \eqref{dr} becomes $-g(\tilde{\varepsilon}_k/g)^2\int dp'\cos(p'L/2)\delta(p')\,\tilde{\rho}_{k,p-p'}=-(\tilde{\varepsilon}_k^2/g)\,\tilde{\rho}_{kp}$, and we get as solution of $\eqref{dr}$ oscillating coherences, as in the hard-wall case, but damped by the probability of the particle to leave the box. The decay time is 
\begin{equation}
 \tau_{d,k}= g/\tilde{\varepsilon}_k^2,
\end{equation}
which expresses the fact that for a harder wall and/or a less energetic incoming particle, the characteristic time to leak out of the box is longer. Slightly penetrable walls corresponds to $g/\tilde{\varepsilon}_k\gg1$.


\bibliographystyle{apsrev}
\bibliography{references}

\end{document}